\documentstyle[preprint,aps]{revtex}
\begin{document}
\draft

\preprint{\it appears in Chemical Physics}

\title{Quantum Effects in Barrier Dynamics}
\author{Joachim Ankerhold and  Hermann Grabert}
\address{Fakult{\"a}t f{\"u}r Physik der
Albert--Ludwigs--Universit{\"a}t,
 Hermann-Herder-Strasse 3,\\ D-79104 Freiburg, Germany}
\date{September 15, 1995}
\maketitle
\begin{abstract}
The dynamics near the top of a potential barrier is studied in the
temperature region where quantum effects become important. The time
evolution of the density matrix of a system that deviates initially
from equilibrium in the vicinity of the barrier top but is in local
equilibrium away from the barrier top is determined. Explicit results
are given for a range of parameters where the nonequilibrium state
is not affected by anharmonicities of the barrier potential except for
the
barrier height. In particular, for a system confined initially to one
side
 of the barrier  the relaxation to a quasi--stationary flux state is
 determined. The associated rate constant is evaluated  and the
relation
 to other rate formulas is discussed in detail.
\end{abstract}
\pacs{PACS numbers:82.20.-w, 05.40.+j}

\narrowtext
\section{Introduction}

The dynamics of systems  hindered by a potential barrier plays an
important
 role in almost all areas of physics and chemistry. The reaction
coordinate
 which describes the transition across the barrier typically interacts
with
 many degrees of freedom. In the classical region, i.e.\ for  high
 temperatures, the generalized Langevin equation for the reaction
coordinate
 usually provides an adequate description of the barrier dynamics.
Based
 on these stochastic methods,   Kramers flux over population approach
 enables a detailed investigation of the escape process across the
 barrier \cite{hanggi2}.

A corresponding formulation of escape processes in the presence of
quantum mechanical effects is available only since recently
\cite{general,dynamic}. In the classical region where the barrier is
crossed by thermally activated processes only the barrier height and
the curvature of the potential at the barrier top and the well minimum
are relevant for the rate constant \cite{hanggi2}. In this article we
study a region where quantum effects lead to large deviations from
classical rate constants but where the harmonic approximation for the
barrier potential is still sufficient to determine the dynamics of the
nonequilibrium state.  This allows for analytical results. We extend
earlier results on quasi--stationary states of systems with large
barriers to include the short time dynamics and the relaxation to a
quasi--stationary flux state. Furthermore, the approach will be used
to determine time correlation functions of population or flux
operators associated with the escape process. This provides the
connection with other familiar rate formulas.

The article is organized as follows. In section \ref{dyn} we give a
brief outline of the formalism and collect results which are of
relevance in the following. In section \ref{escape} the time evolution
of an initial state which is in nonequilibrium near the top of the
barrier is investigated. In section \ref{decay} these results are
applied to determine expectation values, as e.g.\ the avarage flux
across the barrier, and the flux-flux correlation function. The
results are illustrated by an explicit example. Finally, the relation
to other approaches of quantum rate theory is discussed.

\section{Dynamics Near the Barrier Top}\label{dyn}

In this section we collect some results on the description of
dissipative systems that are needed in the following sections and
introduce basic notation.

\subsection{Dynamics of dissipative systems}

The stochastic motion of a classical particle of mass $M$ moving in a
potential field $V(q)$ coupled to a heat bath environment is described
by the generalized Langevin equation
\begin{equation}
M\ddot{q}(t)+M \int_0^t\! {\rm d}t' \, \gamma(t-t') \dot{q}(t') +
\frac{{\rm d}^2 V(q)}{{\rm d} q^2}= \xi(t).\label{eq:nr1}
\end{equation} 
Here, the stochastic force $\xi(t)$ and the nonlocal damping kernel
$\gamma(t-t')$ are connected by the fluctuation--dissipation theorem
\begin{equation}
\langle \xi(t) \xi(t')\rangle= k_{\rm B} T M \gamma(|t-t'|)\label{eq:nr2}
\end{equation}
where $T$ is the temperature of the environment and $k_{\rm B}$
denotes the Boltzman constant.  In this paper we consider systems
where $V(q)$ has a smooth potential barrier. Then, near the barrier
top the barrier potential can be approximated by the potential of an
inverted harmonic oscillator.  Assuming that the barrier top is at
$q=0$ and $V(0)=0$, the barrier potential may be written as
\begin{equation}
V(q)= -\frac{1}{2} M \omega_0^2 q^2.  \label{eq:a1}
\end{equation}
Within the range of coordinates where this form of the potential is
valid, the classical barrier dynamics can be determined exactly by
means of the Langevin equation (\ref{eq:nr1}).  In particular, the
dynamics near the barrier top depends on local features of the barrier
potential only and is not affected by anharmonicities. However, when
the temperatures is lowered quantum effects become important and the
barrier dynamics may depend on global features of the potential field.

The dynamics of a quantum statistical system is determined by the time
evolution of the corresponding density matrix.  Starting at $t=0$ from
a general initial state $W_0$ of the entire system composed of the
Brownian particle and the heat bath, one has
\begin{equation}
W(t) = \exp(-i H t/\hbar) W_{\rm 0} \exp(i H t/\hbar)\label{eq:a3}
\end{equation}
where $H$ contains the Hamiltonians of the system, the environmental
degrees of freedom, and a system-environment coupling.  We shall
assume that the state $W_0$ is out of thermal equilibrium due to a
preparation affecting the degrees of freedom of the Brownian particle
only.  Since we are interested in the dynamics of the particle only,
the time evolution of the reduced density matrix $\rho(t)= {\rm tr_R}
W(t)$ will be considered, where ${\rm tr_R}$ is the trace over the
reservoir. To eliminate the environmental degrees of freedom it is
convenient to employ the path integral approach
\cite{feynman,schulman}. The environmental degrees of freedom can be
integrated out exactly if the heat bath consists of harmonic
oscillators which are coupled linearely to the coordinate of the
particle. In the limit of infinitly many bath oscillators with a
continous frequency spectrum this model causes dissipation and in the
classical limit the generalized Langevin equation (\ref{eq:nr1}) is
recovered.  The details of the path integral representation of the
reduced density matrix and explicit calculations are given elsewhere
\cite{report}. As a result, the position representation of the time
dependent reduced density matrix is found to read
\begin{equation}
\rho(q_f,q'_f,t) 
  = \int\! {\rm d}q_i {\rm d}q'_i {\rm d}{\bar{q}\,}\,{\rm
d}{\bar{q}\,}^\prime\, \,
J(q_f,q_f,t,q_i,q_i,{\bar{q}\,},{\bar{q}\,}^\prime)
\:\lambda(q_i,q_i,{\bar{q}\,},{\bar{q}\,}^\prime). \label{eq:a5}
\end{equation}
Here, $J(q_f,q_f,t,q_i,q_i,{\bar{q}\,},{\bar{q}\,}^\prime)$ denotes
the propagating function represented as a 3-fold path integral where
two path integrals are in real time arising from the two
time--dependent operators in (\ref{eq:a3}) and one in imaginary time
describes system--bath correlations in the initial state. Since for
the parabolic barrier (\ref{eq:a1}) the propagating function is given
explicitly below, we omit here its general form and refer to
\cite{report}.  Equation (\ref{eq:a5}) determines the time evolution
of the density matrix starting from the initial state
\begin{equation}
\rho(q_f,q'_f,0) =  \int\! {\rm d}{\bar{q}\,}\,{\rm
d}{\bar{q}\,}^\prime\,
 \lambda(q_f,{\bar{q}\,},q'_f,{\bar{q}\,}^\prime)
 \rho_\beta({\bar{q}\,},{\bar{q}\,}^\prime) , \label{eq:r19b0}
\end{equation}
where $\rho_\beta = {\rm tr_R}(W_\beta)$. Here, $W_\beta$ is the
equilibrium density matrix of the entire system and
$\lambda(q_f,{\bar{q}\,},q'_f,{\bar{q}\,}^\prime)$ is a preparation
function describing the deviation from thermal equilibrium.  In an
initial state of the form (\ref{eq:r19b0}) the system and the bath are
correlated.  Hence, the customary assumption that the initial density
matrix $W_0$ factorizes into the density matrix of the particle and
the canonical density matrix of the unperturbed heat bath is
avoided. This is a crucial point since (\ref{eq:a5}) allows for the
investigation of the dynamics of realistic physical systems also for
short times where preparation effects are important.

\subsection{Reduced density matrix for an inverted harmonic
oscillator}
\label{reduced}

 In \cite{general} we have shown that anharmonicities of the barrier
potential are always essentiell for very low temperatures. Here, we
investigate the region of high to intermediate temperatures where the
parabolic approximation (\ref{eq:a1}) for the barrier potential is
sufficient but quantum effects may be important.

For the harmonic potential (\ref{eq:a1}) the path integrals involved
in the propagating function can be solved exactly. The explicit
calculation is performed in \cite{general}. One finds
\begin{equation}
\rho(x_f,r_f,t) = \int\! {\rm d}x_i\, {\rm d}r_i\, {\rm d}\bar{x}\,
 {\rm d}\bar{r}\,   J(x_f,r_f,t,x_i,r_i,\bar{x},\bar{r}) \:
\lambda(x_i,r_i,\bar{x},\bar{r}) \label{eq:a9} 
\end{equation}
where we have introduced sum and difference coordinates
\begin{equation}
\begin{array}{ll}
x = q- q{^\prime}, & r = (q + q{^\prime})/2 \label{eq:a8}
\end{array}
\end{equation}
for $q_f$, $q'_f$ and $q_i$, $q'_i$ as well as for $\bar{q}$,$
{\bar{q}\,}^\prime$, respectively.  For the propagating function one
obtains
\begin{eqnarray}
J(x_f,r_f,t,x_i,r_i,\bar{x},\bar{r}) &=& \frac{1}{Z} \,\frac{1}{4 \pi
|A(t)|}\,\frac{1}{\sqrt{\omega_0^2
\hbar\beta|\Lambda|}}\sqrt{\frac{M}{2\pi \hbar^2\beta}}\,
\left(\prod_{n=1}^{\infty} \nu_n^2\, u_n\right) \nonumber \\ & &
\nonumber \\ & & \mbox{}\!\! \times {\exp}\left(\frac{i}{\hbar}
\Sigma_\beta(\bar{x},\bar{r}) +
\frac{i}{\hbar}\Sigma_t(x_f,r_f,t,x_i,r_i,\bar{x},\bar{r})
\right). \nonumber \\ & & \label{eq:a10}
\end{eqnarray}
Here,
\begin{equation}
 \Sigma_\beta(\bar{x},\bar{r}) = i \frac{M }{2 \Lambda}\bar{r}^2 + i
\frac{M \Omega}{2} \, \bar{x}^2
\label{eq:a11}
\end{equation}
is the well-known minimal imaginary-time action of a damped inverted
harmonic oscillator at inverse temperature $\beta=1/k_{\rm B} T$ where
\begin{equation}
\Lambda = \frac{1}{\hbar\beta} \sum_{n=-\infty}^{\infty}\, u_n \label{eq:a12}
\end{equation}
and
\begin{equation}
\Omega = \frac{1}{\hbar\beta} \sum_{n=-\infty}^{\infty} 
 \left(                |\nu_n| \hat{\gamma}(|\nu_n|) - \omega_0^2\right) u_n.
 \label{eq:a13}
\end{equation}
Furthermore,
\begin{equation}
\nu_n = \frac{2 \pi n}{\hbar\beta} \label{eq:a14}
\end{equation}
are Matsubara frequencies and
\begin{equation}
u_n = \left( \nu_n^2 + |\nu_n|\hat{\gamma}(|\nu_n|)
-\omega_0^2\right)^{-1}.\label{eq:a14a}
\end{equation}
 $\hat{\gamma}(z)$ denotes the Laplace transform of the macroscopic
damping kernel $\gamma(s)$ which is determined by the spectral density
$I(\omega)$ of the heat bath
\begin{equation}
\gamma(s) = \frac{2}{M} \int_{0}^{\infty}\frac{{\rm d}{\omega}}{\pi}
\frac{{I}({\omega})}{{\omega}} \cos({\omega} s). \label{eq:a15} 
\end{equation}
We note that for a harmonic oscillator the functions $\Lambda$ and
$\Omega$ correspond to the variance of the position and of the
momentum, respectively. However, for a barrier there is no obvious
physical meaning since e.g.\ for high temperatures one has
$\Lambda<0$. When the temperature is lowered $|\Lambda|$ becomes
smaller and vanishes for the first time at a critical temperature
$T_c$. As seen from (\ref{eq:a10}) and (\ref{eq:a11}) this leads to a
divergence of the propagating function. Hence, as already discussed in
\cite{general}, the harmonic approximation is limited to temperatures
above the critical temperature $T_c$. For temperatures near and below
$T_c$ anharmonicities of the barrier potential field are always
essential \cite{dynamic}.

Apart from the pre--exponential factor the time dependence of the
 propagating function is contained in the second part of the exponent
 of (\ref{eq:a10}). One finds \cite{general} \begin{eqnarray}
\lefteqn{\Sigma_t(x_f,r_f,t,x_i,r_i,\bar{x},\bar{r}) =} \nonumber \\
   & &\left( x_f r_f + x_i r_i \right)M \frac{\dot{A}(t)}{A(t)} + x_i
r_f \frac{\hbar}{2 A(t)}- x_f r_i \frac{2}{\hbar}M^2 \left(\ddot{A}(t)
- \frac{\dot{A}(t)^2}{A(t)}\right) \nonumber \\ & &+ \bar{r} \, x_i M
\left(-\frac{\dot{A}(t)}{A(t)}-\frac{{S}}{2\Lambda A(t)}\right) +
\bar{r} \, x_f\frac{M^2}{\hbar} \left[ 2\left( \ddot{A}(t) -
\frac{\dot{A}(t)^2}{A(t)}\right) + \frac{\dot{S}}{\Lambda} -
\frac{{S}}{\Lambda}\frac{\dot{A}(t)}{A(t)}\right]
\nonumber \\
& &+ i \bar{x} x_i M \left( -\Omega + \frac{\dot{S}}{2 A(t)}\right) -
i \bar{x} x_f\frac{M^2}{\hbar} \left( \ddot{S}(t)
-\frac{\dot{A}(t)}{A(t)} \dot{S}(t) \right) \nonumber \\ & &+
\frac{i}{2} x_i^2 M \left[\Omega-\frac{\dot{S}}{ A(t)} +
\frac{\hbar^2\Lambda}{4 M^2 A(t)^2} \left(1 - \frac{M^2
S(t)^2}{\hbar^2\Lambda^2} \right)\right] \nonumber \\ & &+ i x_i
x_f\frac{M^2}{\hbar} \left[ \ddot{S}(t) -\frac{\dot{A}(t)}{A(t)}
\dot{S}(t)- \frac{\hbar^2\Lambda}{2 M^2 A(t)^2} \left\{ \dot{A}(t)
\left(\frac{M^2 S(t)^2}{\hbar^2 \Lambda^2} - 1 \right) -A(t)\frac{S(t)
\dot{S}(t)M^2}{\Lambda^2\hbar^2}\right\}\right]\nonumber \\ & &+
\frac{i}{2} x_f^2 M \left[ \Omega + \Lambda
\frac{\dot{A}(t)^2}{A(t)^2}- \frac{M^2}{\hbar^2\Lambda}
\left(\dot{S}(t)
-\frac{\dot{A}(t)}{A(t)}S(t)\right)^2\right]. \label{eq:a16}
\end{eqnarray}
Hence, the dynamics at a parabolic barrier is essentially determined
by the functions $A(t)$ and $S(t)$. They are given by the Laplace
transforms of\cite{report}
\begin{equation}
\hat{A}(z) = -\frac{\hbar}{2M}\left(
 z^2 + z \hat{\gamma}(z) -\omega_0^2\right)^{-1}. \label{eq:a17}
\end{equation}
and
\begin{equation}
\hat{S}(z) = \frac{2}{\hbar\beta} \sum_{n=-\infty}^{\infty}
 \frac{z}{z^2-\nu_n^2} \left(\hat{A}(z)-\hat{A}(|\nu_n|)\right).
\label{eq:a18}
\end{equation}

Within the harmonic approximation the above formulas
(\ref{eq:a9})--(\ref{eq:a18}) determine the time evolution of the
density matrix near the top of a potential barrier starting from an
initial state with a deviation from thermal equilibrium described by
the preparation function $\lambda(x_i,r_i,\bar{x},\bar{r})$.

\section{Dynamics of the Escape Process}\label{escape}

Now, we consider a system in a metastable state which may decay by
crossing a potential barrier. We imagine that the system starts out
from a potential well to the left of the barrier.  Metastability means
that the barrier height $V_b$ is much larger than other relevant
energy scales of the system such as $k_{\rm B} T$ and $\hbar
\omega_0$, where $\hbar \omega_0$ is the excitation energy in the well
of the inverted potential. In the temperature region where
anharmonicities can be neglected, i.e.\ for temperatures sufficiently
above $T_c$, the time evolution of an initial nonequilibrium state
near the barrier top can be calculated with the propagating function
(\ref{eq:a10}). In particular, for a system prepared at $t=0$ in
thermal equilibrium in the metastable well, the relaxation to the
quasi--stationary state with constant flux across the barrier can be
investigated. This will be done in this section. The stationary flux
state was already determined in \cite{general} by evaluating the
propagating function in the large time limit.  These investigations
are extendend in the following to include the short time dynamics and
the relaxation to the quasi--stationary state.  Firstly, in
\ref{initial} we introduce the initial preparation. Then, in
\ref{time} we determine the time dependent density matrix, and in
\ref{relaxation} the relaxation to stationary nonequilibrium state is
investigated.

\subsection{Initial preparation}\label{initial}

The initial nonequilibrium state at time $t=0$ is described by the
preparation function \cite{general}
\begin{equation}
\lambda(x_i,r_i,\bar{x},\bar{r}) = \delta(x_i-\bar{x})
\delta(r_i-\bar{r})
 \Theta(-r_i) \label{eq:b1}
\end{equation} 
so that the initial state is a thermal equilibrium state restricted to
the left side of the barrier only. Then, according to (\ref{eq:a9}),
the dynamics is given by
\begin{equation}
\rho(x_f,r_f,t) = \int\! {\rm d}x_i\, {\rm d}r_i\, 
  \tilde{J}(x_f,r_f,t,x_i,r_i) \:\Theta(-r_i) \label{eq:b2} 
\end{equation}
with
\begin{equation}
\tilde{J}(x_f,r_f,t,x_i,r_i)= J(x_f,r_f,t,x_i,r_i,x_i,r_i).\label{eq:b2a}
\end{equation}
In this case the time dependent part of the exponent in the
 propagating function (\ref{eq:b2a}) simplifies to read
\begin{eqnarray}
\lefteqn{\tilde{\Sigma}_t(x_f,r_f,t,x_i,r_i) =
\Sigma_t(x_f,r_f,t,x_i,r_i,x_i,r_i)=} \nonumber \\
   & &\makebox[0.25in][l]{ }x_f r_f M \frac{\dot{A}(t)}{A(t)} + x_i
r_f \frac{\hbar}{2 A(t)} - r_i x_i \frac{{M S(t)}}{2\Lambda A(t)} +
r_i x_f\frac{M^2}{\hbar} \left( \frac{\dot{S}(t)}{\Lambda} -
\frac{{S}(t)}{\Lambda}\frac{\dot{A}(t)}{A(t)}\right)
\nonumber \\
& &\makebox[0.25in][l]{ }+ \frac{i}{2} x_i^2 M \left[-\Omega+
\frac{\hbar^2\Lambda}{4 M^2 A(t)^2} \left(1 - \frac{M^2
S(t)^2}{\hbar^2\Lambda^2} \right)\right] \nonumber \\ &
&\makebox[0.25in][l]{ }- i x_i x_f\frac{\hbar\Lambda}{2 A(t)^2} \left[
\dot{A}(t) \left(\frac{M^2 S(t)^2}{\hbar^2\Lambda^2} - 1 \right)
-A(t)\frac{S(t) \dot{S}(t)M^2}{\Lambda^2\hbar^2}\right]\nonumber \\ &
&\makebox[0.25in][l]{ }+ \frac{i}{2} x_f^2 M \left[ \Omega + \Lambda
\frac{\dot{A}(t)^2}{A(t)^2}- \frac{M^2}{\hbar^2\Lambda}
\left(\dot{S}(t)
-\frac{\dot{A}(t)}{A(t)}S(t)\right)^2\right]. \label{eq:b4}
\end{eqnarray}

\subsection{Time dependent density matrix}\label{time}

Since the exponents (\ref{eq:a11}) and (\ref{eq:b4}) in the
propagating function are bilinear functions of the coordinates, the
integrals in (\ref{eq:b2}) are Gaussian and can be evaluated
exactly. For large times this calculation is performed in detail in
\cite{general}. For arbitrary times we may proceed accordingly.  After
determining the extremum of the exponent in the propagating function
(\ref{eq:b2}) with respect to $x_i$ and $r_i$, one first evaluates the
$x_i$--integral. Then, after simple manipulations of the remaining
$r_i$--integral, the time dependent density matrix may be written in
the form
\begin{equation}
\rho(x_f,r_f,t) = \rho_\beta(x_f,r_f) \, g(x_f,r_f,t).  \label{eq:b17}
\end{equation}
Here,
\begin{equation}
\rho_\beta(x,r) = \frac{1}{Z}
\frac{1}{\sqrt{\omega_0^2\hbar\beta|\Lambda|}}
\,  \sqrt{\frac{M}{2 \pi\hbar^2\beta}} \,\left(\prod_{n=1}^{\infty}
 \nu_n^2\,  u_n\right)\ \exp\left(\frac{i}{\hbar}
\Sigma_\beta(x,r)\right)
 \label{eq:b11}
\end{equation}
is the equilibrium density matrix for an inverted harmonic oscillator
and
\begin{eqnarray}
g(x,r,t)&=&\frac{1}{\sqrt{\pi}}\, \int_{-\infty}^{u(x,r,t)} \!{\rm d}
z \, \exp\left( - z^2\right)\nonumber\\ &=&\frac{1}{2} {\rm
erfc}\left[-u(x,r,t)\right]\label{eq:bg1}
\end{eqnarray}
is a form factor describing deviations from equilibrium with
\begin{equation}
 u(x,r,t)=
\sqrt{\frac{M}{2\hbar|\Lambda|}}\left(1-\frac{\hbar^2\Lambda^2}{M^2
S(t)^2}\right)^{-1/2}\, \left(- r + i |\Lambda|\,
\frac{\dot{S}(t)}{S(t)}\, x\right).\label{eq:b18}
\end{equation}
Clearly, the harmonic approximation is valid only for high enough
temperatures. For temperatures near the critical temperature $T_c$
where $|\Lambda|$ vanishes, the above result becomes divergent.

\subsection{Relaxation to stationary nonequilibrium state}\label{relaxation}

Now, we investigate the dynamics of the density matrix (\ref{eq:b17})
starting from the initial state at $t=0$ in greater detail. Note that
the time dependence of the form factor (\ref{eq:b17}) is completely
determined by the function $S(t)$.

Firstly, let us consider small times $\omega_0 t\ll 1$. There, one has
\cite{report}
\begin{equation}
S(t) = \frac{\hbar\Lambda}{M}-\frac{\hbar\Omega}{2M} t^2 +{\cal
O}(t^4)\label{eq:b19}
\end{equation}
which leads to
\begin{equation}
1-\frac{\hbar^2\Lambda^2}{M^2 S(t)^2} = \frac{\Omega}{|\Lambda|} t^2 +
{\cal O}(t^3).\label{eq:b20}
\end{equation}
Then, the function $u(x,r,t)$, which gives the upper bound of
integration in (\ref{eq:bg1}), reads
\begin{equation}
u(x,r,t)= - r \sqrt{\frac{M}{2\hbar\Omega}}\, \frac{1}{t}+i x
\sqrt{\frac{M \Omega}{2\hbar}}+{\cal O}(t).\label{eq:b21}
\end{equation}
Hence, using the asymptotic formula
\begin{equation}
\int_z^\infty {\rm d}x  \exp(-x^2) \simeq \frac{1}{2 z} \exp(-z^2)
 \ \ \ \mbox{      for      }\mbox{Re}\{z\}\to \infty \label{eq:b22}
\end{equation}
 where Re denotes the real part, the leading order expression for the
form factor (\ref{eq:bg1}) in the limit $\omega_0 t\ll 1$ is found to
read for finite $r$
\begin{equation}
 g(x,r,t)= \Theta(-r) + \sqrt{\frac{\hbar\Omega}{2 M
\pi}}\frac{t}{r}\exp\left( - \frac{M r^2}{2\hbar\Omega t^2} + i
\frac{M x r}{\hbar t} + \frac{M\Omega}{2\hbar}
x^2\right)\label{eq:b23}
\end{equation}
while for $r=0$
\begin{equation}
 g(x,0,t)= \frac{1}{2} + \frac{1}{\sqrt{\pi}}\int_0^{i x
\sqrt{M\Omega/2\hbar}}\!\!\!{\rm d}z\, \exp(-z^2)+{\cal
O}(t).\label{eq:b23b}
\end{equation}
Clearly, for $t\to 0+$ and $r\neq 0$ the form factor reduces to the
$\Theta$ function contained in the initial preparation (\ref{eq:b1})
as expected. On the other hand, at $r=0$ the $t\to 0+$ limit differs
from the $t\to 0-$ limit by an imaginary part due to the discontinuity
of the $\Theta$ function. Defining the width $\Delta(t)$ in position
space of the nonequilibrium state (\ref{eq:b17}) as that value of
$|q|$, $q<0$ where $u(0,q,t)=1$, one gets
\begin{equation}
\Delta(t) =  \sqrt{\frac{2 \hbar|\Lambda|}{M}}\ 
 \left(1-\frac{\hbar^2\Lambda^2}{M^2 S(t)^2}\right)^{1/2}.\label{eq:b24}
\end{equation}
This reduces to $\Delta(t)= \sqrt{2\hbar\Omega/M} t$ for small times
in accordance with (\ref{eq:b23}).

In \cite{general} we have shown that for large times the time
evolution of the density matrix near the barrier top has a stationary
solution. Here, we regain this result from (\ref{eq:b17}).  Evaluating
the functions $A(t)$ and $S(t)$ for times larger than $1/\omega_{\rm
R}$ one gets to leading order an exponential growth \cite{report}
according to
\begin{equation}
A(t)=- \frac{\hbar}{2M}\, \frac{1}{2 \omega_{\rm R} +
\hat{\gamma}(\omega_{\rm R}) + \omega_{\rm R} \hat{\gamma}^\prime
(\omega_{\rm R})} \, \exp( \omega_{\rm R} t) .\label{eq:b25a}
\end{equation} 
and
\begin{equation}
S(t) = - \frac{\hbar}{2M}\, \cot (\frac{\omega_{\rm R}
\hbar\beta}{2})\, \frac{1}{2 \omega_{\rm R} + \hat{\gamma}(\omega_{\rm
R}) + \omega_{\rm R} \hat{\gamma}^\prime (\omega_{\rm R})} \, \exp(
\omega_{\rm R} t) .\label{eq:b25}
\end{equation} 
 Here, $\hat{\gamma}^\prime (z)$ denotes the derivative of
$\hat{\gamma}(z)$, and $\omega_{\rm R}$ is the Grote-Hynes frequency
\cite{grote} given by the positive solution of $\omega_{\rm R}^2 +
\omega_{\rm R} \hat{\gamma}(\omega_{\rm R}) = \omega_0^2$. Eqs.\
(\ref{eq:b25a}) and (\ref{eq:b25}) describe the unbounded motion at
the parabolic barrier with corrections that are exponentially decaying
in time (see \cite{report} for details).  Hence, the function
$u(x,r,t)$ in (\ref{eq:b18}) becomes independent of time
\begin{equation}
 u_\infty = \sqrt{\frac{M}{2\hbar|\Lambda|}}\left(- r + i |\Lambda|\,
\omega_{\rm R} \, x\right), \label{eq:b26}
\end{equation}
and the density matrix (\ref{eq:b17}) reduces to the stationary
nonequilibrium state derived in \cite{general}. This time independent
state describes a constant flux across the potential barrier and
generalizes the well--known Kramers flux state to the temperature
region where quantum effects are important.  The width $\Delta(t)$
from (\ref{eq:b24}) saturates for large times at the finite value
\begin{equation}
\Delta_\infty=  \sqrt{\frac{2\hbar|\Lambda|}{M}}\label{eq:b27}
\end{equation}
which coincides with the width of the diagonal part of the equilibrium
distribution (\ref{eq:b11}).

From the above discussion it is obvious that a lower bound of time
 where the stationary flux solution holds derives from $\omega_{\rm R}
 t\gg 1$. For very long times depletion of states inside the potential
 well leads to a flux decreasing in time. Hence, for very long times
 anharmonicities of the barrier potential become important. For a
 barrier potential with a quartic term as leading order anharmonicity
 the upper bound of time where the density matrix (\ref{eq:b17}) is
 valid has been estimated in \cite{general}. One obtains the condition
 $\exp(\omega_{\rm R} t)\ll q_a \sqrt{2M\omega_0/\hbar|\Lambda|}$
 where $q_a$ denotes a characteristic length indicating a typical
 distance from the barrier top at which the anharmonic part of the
 potential becomes essentiell.

 The density matrix (\ref{eq:b17}) depends on local properties of the
metastable potential near the barrier top only. On the other hand, the
metastable state is assumed to be in thermal equilibrium near the well
bottom. This means that the solution (\ref{eq:b17}) must reduce to the
thermal equilibrium state for coordinates $q_f$, $q_f^\prime$ on the
left side of the barrier at distances small compared with $q_a$. Now,
for $t=0$ the equilibrium state extends to the top of the barrier and
the matching to the equilibrium state in the well is most critical for
the stationary flux state where $\Delta(t)$ is largest. However, this
latter case was examined in \cite{general}. One obtains the condition
\begin{equation}
|\Lambda| \ll \frac{V_b}{\hbar\omega_0^2}\left( 1-\omega_{\rm R}^2
\frac{|\Lambda|}{\Omega}\right)\label{eq:ca8}
\end{equation}
where $V_b$ is the barrier height with respect to the well bottom.
From a physical point of view (\ref{eq:ca8}) defines the region where
the influence of the heat bath on the escape dynamics is strong enough
to equilibrate particles on a length scale smaller than the scale
where anharmonicities becomes important. Only then nonequilibrium
effects remain localized in coordinate space to the barrier region
also for longer times. Especially in the classical region where
$k_{\rm B} T\gg \hbar\omega_0$ and for Ohmic damping
$\hat{\gamma}(z)=\gamma$ Eq.\ (\ref{eq:ca8}) reduces to the well-known
Kramers condition \cite{hanggi2} $k_{\rm B} T \omega_0/V_b\ll
\gamma$. Here, $1-\omega_{\rm R}^2\approx \gamma$ for small damping
has been used. When the temperature is lowered $|\Lambda|$ decreases
and the range of damping where the stationary solution (\ref{eq:b17})
is valid becomes larger. This is investigated in detail in
\cite{general}.

\section{Decay Rate and Relation to other Approaches}\label{decay}

In this section the time dependent density matrix derived above is
used to evaluate expectation values, in particular the average flux
across the barrier. Further, the relation of the theory to other
approaches to rate constants is discussed.

\subsection{Average flux and decay rate}

Clearly, the solution (\ref{eq:b17}) contains all relevant information
about the nonequilibrium state. Now, we want to evaluate the total
probability flux at the barrier top $q=0$.  One has
\begin{equation}
J(t) = \frac{1}{2 M} \langle \hat{p} \delta(\hat{q}) + \delta(\hat{q})
\hat{p}\rangle_t\label{eq:d1}
\end{equation}
where the expectation value $\langle \cdot\rangle_t$ is calculated
  with respect to the time dependent nonequilibrium state.  From
  (\ref{eq:d1}) one has in coordinate representation
\begin{equation}
J(t) = \left.  \frac{\hbar}{iM} \frac{\partial}{\partial x_f} \,
\rho(x_f,0,t)\right |_{x_f=0}.\label{eq:d2}
\end{equation}
Since the essential contribution to the population in the well comes
from the region near the well bottom, the normalization constant $Z$
in (\ref{eq:d2}) can be approximated by the partition function of a
damped harmonic oscillator with frequeny $\omega_{\rm w}$ at the well
bottom, i.e.
\begin{equation}
Z = \frac{1}{\omega_{\rm w} \hbar \beta} \left(\prod_{n=1}^{\infty}
\frac{\nu_n^2}{\nu_n^2 + |\nu_n| {\hat{\gamma}(|\nu_n|)} + \omega_{\rm
w}^2}\right) \exp(\beta V_b). \label{eq:d3}
\end{equation}
Here, $V_b$ denotes the barrier height with respect to the well
bottom. Note that the potential was set to 0 at the barrier top.
Inserting (\ref{eq:b17}) for $r_f=0$ and (\ref{eq:d3}) into
(\ref{eq:d2}) one obtains
\begin{equation}
J(t)= \Gamma\ \eta(t)\label{eq:d4}
\end{equation}
where
\begin{eqnarray}
\Gamma & =&\lim_{t \to \infty} J(t)\nonumber\\
&= & \frac{\omega_{\rm w}}{2 \pi} \, \omega_{\rm R} \, \left(
\prod_{n=1}^{\infty} \frac{\nu_n^2 + |\nu_n| \hat{\gamma}(|\nu_n|) +
\omega_{\rm w}^2}{\nu_n^2 + |\nu_n| \hat{\gamma}(|\nu_n|)
-\omega_0^2}\right)\, \exp(- \beta V_b) \label{eq:d5}
\end{eqnarray}
denotes the decay rate of the metastable system in the well. We recall
that the Grote-Hynes frequency $\omega_{\rm R}$ is given by the
positive solution of $\omega_{\rm R}^2 + \omega_{\rm R}
\hat{\gamma}(\omega_{\rm R}) = \omega_0^2$. The rate (\ref{eq:d5})
describes thermally activated transitions across the barrier where the
prefactor takes into account quantum corrections
\cite{general,grabert-olschowski,wolynes}.  For the time dependent
function $\eta(t)$ one gets
\begin{equation}
\eta(t)=\frac{\dot{S}(t)}{\omega_{\rm R}\, S(t)}\,
 \left( 1- \frac{\hbar^2\Lambda^2}{M^2 S(t)^2}\right)^{-1/2}.\label{eq:d6}
\end{equation}
This way we have found an analytical result for the dynamic behavior
of the average flux which is usually studied numerically, see e.g.\
\cite{wolynes2}.  For long times $\omega_{\rm R}t\gg 1$ the above
function approaches 1. For very small times one obtains from
(\ref{eq:b19})
\begin{equation}
\eta(t)=\frac{1}{\omega_{\rm R}}\,
\sqrt{\frac{\Omega}{\omega_0^2|\Lambda|}}
+ {\cal O}(t^2)\label{eq:d7}
\end{equation}
which gives a finite flux for $t\to 0+$ while, according to the
initial preparation (\ref{eq:b1}), the limit $t\to 0-$ leads to a
vanishing flux [see also (\ref{eq:b23}) and
(\ref{eq:b23b})]. Specifically, for finite damping
\begin{equation}
\eta(0) = \frac{1}{\omega_{\rm R}}\,
 \sqrt{\frac{\Omega}{\omega_0^2|\Lambda|}}\label{eq:d8}
\end{equation}
is always larger than 1. As a consequence, the probability flux for
$t\to 0+$ exceeds the rate (\ref{eq:d5}). For very high temperatures
where $\hbar\beta\ll 1$, Eq.\ (\ref{eq:d8}) reduces to
$\eta(0)=1/\omega_{\rm R}$. The corresponding probability flux
$J(0)=\Gamma/\omega_{\rm R}$ coincides with the result of classical
transition state theory \cite{hanggi2}
\begin{equation}
\Gamma_{\rm cl}=  \frac{\omega_{\rm w}}{2\pi} \, \exp(- \beta V_b).
 \label{eq:d9}
\end{equation}
Here, we have used the fact that the term in brackets in the prefactor
of (\ref{eq:d5}) approaches 1 for $\hbar\beta\ll 1$. For lower
temperatures $|\Lambda|$ decreases and $\eta(0)$ becomes larger than
$1/\omega_{\rm R}$.

\subsection{Flux--flux correlation function}

The propagating function can also be used to determine correlation
functions. Here we consider the right--left spatial correlation
function
\begin{equation}
C_{\rm R L}(t)= {\rm tr} \left\{ \Theta[q(t)] \Theta[-q]
\rho_\beta\right\}= \langle \Theta[q(t)]\,
\Theta[-q]\rangle_\beta\label{eq:ff1}
\end{equation}
where $\Theta(\cdot)$ denotes the step function. Time derivatives of
$C_{\rm R L}(t)$ lead to further correlation functions, in particular
the flux--flux correlation. Below we will see that these correlations
are connected with other rate formulas.

Now, let us evaluate $C_{\rm R L}(t)$ explicitly. Within the presented
real time approach this correlation function may formally be looked
upon as the expectation value of $\Theta(q)$ at time $t$ of a system
with an initial ``density matrix'' $\Theta(-q) \rho_\beta$. The
corresponding preparation function then takes the form
\begin{equation}
\lambda(x_i,r_i,\bar{x},\bar{r})= \Theta\left(-r_i-x_i/2\right)\,
 \delta(x_i-\bar{x})\, \delta(r_i-\bar{r}).\label{eq:lamf}
\end{equation}
This way, using (\ref{eq:a9}), the correlation function may be written
as
\begin{eqnarray}
C_{\rm R L}(t)&=& \int {\rm d} r_f {\rm d}x_i {\rm d}r_i\, \Theta(r_f)
\Theta\left(-r_i-x_i/2\right) \tilde{J}(0,r_f,t,x_i,r_i)\nonumber\\
&=& \int {\rm d} r_f {\rm d}x_i {\rm d}r_i'\, \Theta(r_f)
\Theta(-r_i') \tilde{J}(0,r_f,t,x_i,r_i'-x_i/2)\label{eq:jf}
\end{eqnarray}
where the propagating function $\tilde{J}(x_f,r_f,t,x_i,r_i)$ is given
in (\ref{eq:b2a}).  We proceed as in section \ref{time} and first
evaluate the $x_i$ and afterwards the $r_i$ integration.  Here, the
maximum of the exponent in the propagating function with respect to
$x_i$ and $r_i'$ lies at
\begin{eqnarray}
x_i^0 &=& i \frac{2M\omega_0}{\hbar} A(t)
\frac{r_f}{\Lambda}\nonumber\\ {r_i'}^0 &=& \frac{M}{\hbar}[S(t)+i
A(t)] \frac{r_f}{\Lambda}.\label{eq:maxf}
\end{eqnarray}
Introducing shifted coordinates $\hat{x}_i=x_i-x_i^0$ and
$\hat{r}_i'=r_i'-{r_i'}^0$ a straightforward calculation shows that
\begin{eqnarray}
\lefteqn{\Sigma_\beta(x_i,r_i'-x_i/2)+
\tilde{\Sigma}(0,r_f,t,x_i,r_i'-x_i/2)=}\nonumber\\
& & - \frac{iM \hat{x}_i^2}{8\Lambda A(t)^2}\left\{\left[S(t)+i
A(t)\right]^2 -\frac{\hbar^2\Lambda^2}{M^2}\right\} + \frac{i M
(\hat{r}_i^\prime)^2}{2\Lambda} - \frac{M \hat{x}_i \hat{r}_i'}{2
\Lambda A(t)} \left[ S(t)+iA(t)\right].\label{eq:sigf}
\end{eqnarray}
The Gaussian integrals with respect to $\hat{x}_i$ and $\hat{r}_i'$
are now readily performed. Finally, after some further manipulations,
we end up with
\begin{eqnarray}
C_{\rm R L}(t)&=& \frac{1}{ Z } \frac{1}{ \pi \hbar\beta}
\,\left(\prod_{n=1}^{\infty} \nu_n^2\, u_n\right)\int_0^\infty\!\!{\rm
d}x \exp(x^2)\int_{x/z(t)}^{\infty}\!\!{\rm d}y\,
\exp(-y^2)\nonumber\\ &=&\frac{1}{ Z } \frac{1}{4 \pi \hbar\beta}
\,\left(\prod_{n=1}^{\infty} \nu_n^2\, u_n\right)
\log\left(\frac{1+z(t)}{1-z(t)}\right)\label{eq:resf}
\end{eqnarray}
where
\begin{equation}
z(t)= \left\{ 1-\frac{\hbar^2\Lambda^2}{M^2
[S(t)+iA(t)]^2}\right\}^{1/2}.\label{eq:zf}
\end{equation}
For $t\to 0$ one has from (\ref{eq:a17})
\begin{equation}
A(t)=-\frac{\hbar}{2M} t+ O(t^3).\label{eq:smalla}
\end{equation}
Hence, $z(t)$ tends to zero and $C_{\rm R L}(t)$ vanishes for $t\to 0$
as expected.  Now, the time derivative of (\ref{eq:resf}) yields
\begin{eqnarray}
\dot{C}_{\rm R L} (t)& =& \langle \bar{F}(t)
 \Theta(-q)\rangle_\beta\nonumber\\
&=& \frac{1}{ Z } \frac{1}{2\pi \hbar\beta}
\,\left(\prod_{n=1}^{\infty} \nu_n^2\, u_n\right)\frac{|\dot{S}(t)|+i
|\dot{A}(t)|}{\left\{[S(t)+i A(t)]^2
-\hbar^2\Lambda^2/M^2\right\}^{1/2}}\label{eq:fmf}
\end{eqnarray}
where
\begin{equation}
\bar{F}= \frac{1}{2} \left[p \delta(q)  + \delta(q)p \right]\label{eq:ff7}
\end{equation}
is the flux operator.  Finally, a second time derivative gives the
flux--flux correlation
\begin{eqnarray}
\ddot{C}_{\rm R L} (t)&=& \langle \bar{F}(t) \bar{F}\rangle_\beta\nonumber\\
&=& \frac{1}{ Z } \frac{1}{2\pi \hbar\beta}\left(\prod_{n=1}^{\infty}
 \nu_n^2\, u_n\right)\nonumber\\ &
 &\mbox{}\times\left\{\frac{|\ddot{S}(t)|+i
 |\ddot{A}(t)|}{\left\{[S(t)+i A(t)]^2
 -\hbar^2\Lambda^2/M^2\right\}^{1/2}}-\frac{[|\dot{S}(t)|+i|\dot{A}(t)|]^2
 [S(t)+iA(t)]}{\left\{[S(t)+i A(t)]^2
 -\hbar^2\Lambda^2/M^2\right\}^{3/2}}\right\}.\label{eq:fmfa}
\end{eqnarray}
The above three correlations are related to the escape rate out of the
metastable well as will be seen in section \ref{other}.

\subsection{An example: Drude damping}

To illustrate the above results we now consider a Drude model with
$\gamma(t)=\gamma\omega_{\rm D} \exp(-\omega_{\rm D} t)$ by way of
example. Clearly, in the limit $\omega_{\rm D}\gg \omega_0, \gamma$
the Drude model behaves like an Ohmic model execpt for very short
times of order $1/\omega_{\rm D}$. The Laplace-transform of
$\gamma(t)$ reads
\begin{equation}
\hat{\gamma}(z)=\gamma\frac{\omega_{\rm D}}{\omega_{\rm D} +z}.\label{eq:f1}
\end{equation}
Then, from (\ref{eq:a12}) and (\ref{eq:a13}) we obtain
\begin{equation}
\Lambda = \frac{1}{\hbar\beta} \sum_{n=-\infty}^{\infty}\,
 \frac{1}{\nu_n^2  +| \nu_n | (\gamma\omega_{\rm D}/\omega_{\rm
D}+|\nu_n|)
 - \omega_0^2}\label{eq:f2}
\end{equation}
and
\begin{equation}
\Omega = \frac{1}{\hbar\beta} \sum_{n=-\infty}^{\infty} \,
 \frac{  |\nu_n|(\gamma\omega_{\rm D}/\omega_{\rm D}+|\nu_n|)
  - \omega_0^2}{\nu_n^2  +| \nu_n |(\gamma\omega_{\rm D}/
\omega_{\rm D}+|\nu_n|) - \omega_0^2}. \label{eq:f3}
\end{equation}

The time dependence of the nonequilibrium state is completely
determined by the function $S(t)$ in (\ref{eq:a18}). Some of the
algebra needed to evaluate $S(t)$ for a Drude model explicitly is
provided in recent work \cite{graberttalk}. We obtain
\begin{equation}
S(t)= \frac{\hbar}{M}\sum_{i=1}^3\,
\left[\frac{c_i}{2}\cot\left(\frac{\lambda_i\hbar\beta}{2}\right)
\exp(\lambda_i
t)\right] -
\zeta(t).\label{eq:f4}
\end{equation} 
Here, $\lambda_i$, $i=1,2,3$ denote the poles of $\hat{A}(z)$ given by
the three solutions of
\begin{equation}
z^3+\omega_{\rm D} z^2 + z (\gamma \omega_{\rm D} -\omega_0^2)
-\omega_{\rm D}=0.\label{eq:f5}
\end{equation}
For the coefficients $c_i$ one has
\begin{eqnarray}
c_1&=&(\lambda_2^2-\lambda_3^2)/\phi\nonumber\\
c_2&=&(\lambda_3^2-\lambda_1^2)/\phi\nonumber\\
c_3&=&(\lambda_1^2-\lambda_2^2)/\phi\label{eq:ci}
\end{eqnarray}
where
\begin{equation}
\phi =(\lambda_1-\lambda_2)\lambda_1\lambda_2+
(\lambda_2-\lambda_3)\lambda_2\lambda_3+
(\lambda_3-\lambda_1)\lambda_1\lambda_3.\label{eq:f6}
\end{equation}
Further, we have introduced the time dependent function
\begin{equation}
\zeta(t)=\frac{\gamma\omega_{\rm D}^2}{\hbar\beta}
\sum_{n=-\infty}^{\infty} \frac{|\nu_n|\, \exp(-|\nu_n|t)}
{(\lambda_1^2-\nu_n^2)(\lambda_2^2-\nu_n^2)(\lambda_3^2-\nu_n^2)}.
\label{eq:f7}
\end{equation}
which can also be written in terms of hypergeometric functions as
\begin{equation}
\zeta(t)=-\frac{1}{\hbar\beta}\sum_{i=1}^{3}\frac{c_i}{\lambda_i}
\left[F(1,\frac{\lambda_i}{\nu};1+\frac{\lambda_i}{\nu};{\rm e}^{-\nu
t})
-F(1,-\frac{\lambda_i}{\nu};1-\frac{\lambda_i}{\nu};{\rm e}^{-\nu
t})\right]
.\label{eq:f8}
\end{equation}

With these results for $\Lambda$, $\Omega$, and $S(t)$ and a Drude
frequency $\omega_{\rm D}=100\omega_0$ we have investigated the time
evolution of the nonequilibrium state numerically.  In Fig.\ 1 the
width $\Delta(t)$ of the nonequilibrium state in position space, given
in (\ref{eq:b24}), is depicted as a function of $t$ for various
temperatures. For high temperatures damping effects are relevant for
intermediate times only while for lower temperatures they are
essentiell for all times. For small times $\Delta(t)$ grows faster for
stronger damping and reaches a larger asymptotic value for large
times. This is due to the quantum mechanical effect that stronger
damping suppresses the fluctuations of the coordinate and therefore
enhances fluctuations of the momentum.

The relaxation of the time dependent flux (\ref{eq:d4}) across the
potential barrier to the time independent decay rate (\ref{eq:d5}) is
determined by the function $\eta(t)$ in (\ref{eq:d6}). In Fig.~2 the
time dependence of $\eta(t)$ is depicted for various temperatures. One
sees that in the region of moderate damping the simple TST result
$\Gamma_{\rm TST}=\Gamma \eta(0)$ for the rate constant gives a
satisfactory estimate of the true rate only for high
temperatures. When the temperature is decreased $\eta(0)$ grows and
depends strongly on the damping strength. Furthermore, for lower
temperatures the average flux across the barrier becomes stationary
faster for stronger damping.

\subsection{Relation to other rate formulas}\label{other}

In the previous section we have calculated the probability flux across
the potential barrier using the time dependent density matrix
(\ref{eq:a9}) with the initial preparation (\ref{eq:b1}). In
particular, we have shown that the flux becomes time independent for
times $\omega_{\rm R} t\gg 1$ leading to the escape rate. Here, we
want to regain the escape rate using rate formulas first introduced by
Yamamoto \cite{yama} and Miller \cite{miller}.  First, let us consider
Yamamoto's rate formula
\begin{equation}
\Gamma = \lim_{t\to \infty} \frac{1}{\hbar\beta}\int_0^{\hbar\beta}
 {\rm d}\lambda \langle \Theta[-q(-i\lambda)] \dot{\Theta}[-q(t)]
\rangle_\beta\label{eq:ff2}
\end{equation}
where the limit is understood as $t\gg 1/\omega_{\rm R}$. Here, the
right hand side can be transformed to read
\begin{equation}
\frac{1}{\hbar\beta}\int_0^{\hbar\beta} {\rm d}\lambda \langle
 \Theta[-q(-i\lambda)] \dot{\Theta}[-q(t)]\rangle_\beta= 
 \frac{i}{\hbar\beta} \langle \left[\Theta[-q(t)], 
 \Theta[-q]\right]\rangle_\beta.\label{eq:ff3}
\end{equation}
On the other hand, taking into account that $\Theta(q)=1-\Theta(-q)$
one has from (\ref{eq:ff1})
\begin{equation}
{\rm Im} \left\{C_{\rm R L}(t)\right\}=- {\rm Im} \left\{C_{\rm L
L}(t)\right\}= \frac{i}{2}\langle \Theta[-q(t)]\,
\Theta[-q]\rangle_\beta.\label{eq:ff4}
\end{equation}
Hence, we get from (\ref{eq:ff3})
\begin{equation}
\Gamma=\frac{2}{\hbar\beta} \lim_{t \to\infty} {\rm Im}
 \left\{C_{\rm R L}(t)\right\}.\label{eq:ff5}
\end{equation}
The result (\ref{eq:resf}) can now be inserted into the above rate
formula. First, from (\ref{eq:b25a}) and (\ref{eq:b25}) one obtains
for times $\omega_{\rm R} t\gg 1$
\begin{equation}
 {\rm Im}\left\{\log\left(\frac{1+z(t)}{1-z(t)}\right) \right\}= 2
\arctan\left[A(t)/S(t)\right].\label{eq:yamf}
\end{equation}
Thus, we obtain from (\ref{eq:resf})
\begin{equation}
\lim_{t\to \infty} {\rm Im}\left\{ C_{\rm R L}(t)\right\}=
\frac{\omega_{\rm R}\hbar\beta}{2} \frac{1}{ Z }  \frac{1}
{2 \pi \hbar\beta} \,\left(\prod_{n=1}^{\infty} \nu_n^2\,  u_n\right)
\label{eq:yamfa}
\end{equation}
which combines with (\ref{eq:ff5}) and the normalization (\ref{eq:d3})
to yield the escape rate (\ref{eq:d5}).

On the other hand, the time derivative $\dot{C}_{\rm R L}(t)$ given in
(\ref{eq:fmf}) determines Miller's rate formula \cite{miller}
\begin{equation}
\Gamma= \lim_{t\to\infty} \dot{C}_{\rm R L}(t). \label{eq:ff8}
\end{equation}

In the long time limit the imaginary part of $\dot{C}_{\rm R L}(t)$
becomes exponentially small and
\begin{equation}
\lim_{t\to\infty}\dot{C}_{\rm R L} (t)= \frac{1}{ Z } 
 \frac{1}{2\pi \hbar\beta} \,\left(\prod_{n=1}^{\infty} \nu_n^2\, 
 u_n\right)\omega_{\rm R}\label{eq:fmm}
\end{equation} 
yields with (\ref{eq:ff8}) again the rate (\ref{eq:d5}).

 We note that for long times the flux-flux autocorrelation function
(\ref{eq:fmfa}) becomes exponentially small. This indicates a constant
flux across the barrier independent of the initial preparation of the
nonequilibrium state in the metastable well.

\section{Conclusions}

Within the path integral approach we have evaluated the time dependent
density matrix of a metastable system in the vicinity of a barrier top
when preparing the system at $t=0$ in thermal equilibrium on the left
side of the barrier only (\ref{eq:b1}). The explicit solution
(\ref{eq:b17}) is valid over a wide range of time excluding very long
times and for high as well as for lower temperatures where quantum
effects become important. The nonequilibrium state approaches an
equilibrium state as one moves away from the barrier top. Condition
(\ref{eq:ca8}) on the damping strength ensures that equilibrium is
reached within the range of validity of the harmonic approximation for
the barrier potential.

   In particular, we have studied the relaxation of the time dependent
nonequilibrium state to the stationary flux state. We found that the
corresponding time dependent normalized flux across the barrier is
decaying in time.  For very high temperatures the initial flux
coincides with the transition state theory rate. For long times the
flux coincides with the stationary decay rate of the metastable state
which was shown to be identical with the well--known rate formula for
thermally activated decay in the presence of quantum
corrections. Furthermore, we have shown that the real time approach
can also be used to evaluate correlation functions which are
encountered in other rate formulas.

 \acknowledgements

The authors would like to thank G.-L.\ Ingold and  E.\ Pollak for
valuable discussions. This work was supported by the
Sonderforschungsbereich 237.

\begin{figure}
\caption{Width $\Delta(t)$ in position space of the nonequilibrium
state
 for high temperatures $\hbar\beta\omega_0=0.05$ (thick lines) and
 lower temperatures $\hbar\beta\omega_0=2.0$ (thin lines) for a Drude
 model with $\omega_{\rm D}/\omega_0=100$. Solid lines indicate
 small damping with $\gamma=0.1$, dashed lines stronger damping
 with $\gamma=3.0$.}\label{fig:delta}
\end{figure}

\begin{figure}
\caption{Time dependence of the average flux across the barrier
 $\eta(t)$ for two temperatures and  a Drude model with
 $\omega_{\rm D}/\omega_0=100$ and with various damping strengths.}
\label{fig:eta}
\end{figure}


\begin{references}

\bibitem{hanggi2} P.\ H{\"a}nggi, P.\ Talkner, and M.\ Borkovec,
 { Rev.\ Mod.\ Phys.} {\bf 62}, 251 (1990) and references therein.  


\bibitem{general} J.\ Ankerhold, H.\ Grabert, and G.-L.\ Ingold,
Phys.\ Rev.\ E. {\bf 51}, 4267 (1995)


\bibitem{dynamic} J.\ Ankerhold and H.\ Grabert, appears in Phys.\
Rev.\ E. {\bf 52}

\bibitem{feynman} R.\ P.\ Feynman and A.\ P.\ Hibbs,
 { Quantum Mechanics and Path Integrals} (McGraw-Hill, New York,
1965);
 R.\ P.\ Feynman, { Statistical Mechanics} (Benjamin, New York, 1972).

\bibitem{schulman} L.\ S.\ Schulman, { Techniques and Applications of
 Path Integrals}  (Wiley, New York, 1981).



\bibitem{caldeira}  A.\ O.\ Caldeira and A.\ J.\ Leggett,
 { Phys.\ Rev.\ Lett.} {\bf 46},  211 (1981); A.\ O.\ Caldeira and
 A.\ J.\ Leggett, { Ann.\  Phys.\  (USA)} {\bf 149},  374 (1983);
 {\bf 153},  445(E) (1984).

\bibitem{weiss} U.\ Weiss, { Quantum Dissipative Systems}
 (World Scientific, Singapore, 1993).

\bibitem{report} H.\ Grabert, P.\ Schramm, and G.-L.\ Ingold,
 { Phys.\ Rep.} {\bf 168},  115 (1988).



\bibitem{grote} R.\ F.\ Grote and J.\  T.\ Hynes, { J.\ Chem.\ Phys.}
 {\bf 73},  2715 (1980); P.\ H{\"a}nggi and F.\ Mojtabai,
 { Phys.\ Rev.\ A} {\bf 26}, 1168 (1982).


\bibitem{grabert-olschowski} H.\ Grabert, P.\ Olschowski, and U.\
Weiss,
 { Phys.\ Rev.} {\bf B 36}, 1931 (1987).

\bibitem{wolynes} P.\ G.\ Wolynes, {  Phys.\ Rev.\ Lett.} {\bf 47},
  968 (1981).

\bibitem{wolynes2} Y.\ Tanimura and P.\ G.\ Wolynes, {J.\ Chem.\
Phys.}
 {\bf 96}, 8485 (1992).


\bibitem{graberttalk} H.\ Grabert, U.\ Weiss, and P.\ Talkner,
 { Z.\ Phys.\ B} {\bf 55}, 87 (1984).


\bibitem{yama} T.\ Yamamoto, J.\ Chem.\ Phys.\ {\bf 33}, 281 (1960).

\bibitem{miller} W.\ H.\ Miller, { J.\ Chem.\ Phys.} {\bf 62},
  1899 (1975); W.\ H.\ Miller, { Adv.\ Chem.\ Phys.} {\bf 25},
  69 (1974);  W.\ H.\ Miller, S.\ D.\ Schwartz, and J.\ W.\ Tromp,
 J.\ Chem.\ Phys.\ {\bf 79}, 4889 (1983).


\end{references}
\end{document}